\documentclass[10pt,a4paper]{article}

\usepackage{times}

\usepackage{epsfig,wrapfig}
\usepackage{epstopdf} 

\usepackage{fancyhdr}

\usepackage[lmargin=2.5cm, rmargin=2.5cm,tmargin=3.50cm,bmargin=3.50cm]{geometry}
\usepackage{indentfirst}
\usepackage{subcaption}

\usepackage{titlesec}
\titleformat{\section}[hang]
  {\centering}{\thesection}{1ex}{\normalsize \textsc}
\titleformat{\subsection}[hang]
  {}{\thesubsection}{1ex}{\normalsize \textit}

\renewcommand{\thesection}{ \normalsize \textnormal{\Roman{section}.}}
\renewcommand{\thesubsection}{\normalsize \textnormal{\textsc{\textit{\Alph{subsection}.}}}}

\def\e{\begin{equation}}
\def\f{\end{equation}}
\def\_#1{{\bf #1}}

\def\.{\cdot}

\begin{document}

\title{\large \textbf{A Single-Layer Meta-Atom Absorber}}
%
\def\affil#1{\begin{itemize} \item[] #1 \end{itemize}}
\author{\normalsize \bfseries \underline{I.~A.~Faniayeu}$^{1}$,  V.~S.~Asadchy$^{1,2}$, T.~A.~Dziarzhauskaya$^{1}$, I.~V.~Semchenko$^{1}$ and S.~A.~Khakhomov$^1$
}
\date{}
\maketitle
\thispagestyle{fancy} 
\vspace{-6ex}
\affil{\begin{center}\normalsize $^1$Department of General Physics, Gomel State University, 246019, Belarus\\
$^2$Department of Radio Science and Engineering, Aalto University\\ P.O.~13000, FI-00076 Aalto, Finland\\
bratya.i@mail.ru
 \end{center}}

\begin{abstract}
\noindent \normalsize
\textbf{\textit{Abstract} \ \ -- \ \
We realized and experimentally tested a conceptually new kind of electrically thin absorbers of electromagnetic waves. The idea is to utilize a single layer of precisely designed meta-atoms. This allows one to design an absorber with unprecedentedly small thickness. The absorber implies absence of a ground plane. High efficiency of the realized structure in the S band is demonstrated. The conceptual idea of the proposed absorber can find many applications especially at optical frequencies.}
\end{abstract}

\section{Introduction}

Recently, a significant amount of works has been devoted to electrically thin totally absorbing layers. There has been proposed a variety of different topologies of absorbers. However, electrically thin absorbers mostly imply the use of a ground plane \cite{1} limiting electromagnetic response of the structure from the opposite side. Such two-layer structures possess asymmetrical properties and do not absorb radiation from the opposite side. Also, the metal-backed absorbers forbid transmission at all frequencies being inconvenient for some applications. At infrared and optical frequencies, it is practically impossible to realize a perfect ground plane except the case of electrically thick layers of photonic crystals. Hence, the idea to design an electrically thin single-layer absorber without ground plane has been attracting a big interest of engineers.

In order to achieve total absorption, one needs to provide equally significant electric and magnetic responses of the structure (for matching wave impedance of the absorber to that of vacuum) and certain amount of dissipative loss (to provide full absorption of incident waves). At microwave and higher frequencies permeability of known bulk materials is equal to unit and necessary magnetic response can be accomplished only through the use of artificial metamaterials. 

We realize a conceptually new kind of a perfect absorber for normally incident electromagnetic waves. It consists of only a single array of meta-atoms (small dipolar inclusions). It allows us to design an absorber with extremely small thickness but still which provides necessary magnetic response. The design does not imply presence of a ground plane. Therefore, the two sides of the absorber are operating. The induced electric and magnetic dipole moments in the inclusions under illumination do not scatter in backward direction but in forward direction they radiate secondary plane waves destructively interfering with incident ones (providing zero transmission) \cite{2}. The equal electric and magnetic properties are achieved due to optimal shape of the inclusions and their certain arrangement in the array \cite{3}. The realized absorber demonstrates highly efficient performance and operates symmetrically for incident waves from the two sides. Although here we realize the concept of an optically thin meta-atom absorber only in the microwave frequency range, it can be applied to the higher frequencies with the use of different certain inclusions.

\section{Perfect absorption in an array of meta-atoms}

Meta-atoms are electromagnetic particles of a sub-wavelength size. A regular planar array of the meta-atoms forms an artificial single-layer metamaterial, so-called “metasurface”. Incident waves induce oscillating electric $\mathbf{p}$ and magnetic $\mathbf{m}$ dipoles in the inclusions of a metasurface. In \cite{2} it was shown that symmetrical perfect absorption operation can be accomplished only in a metasurface which has equally significant co-polarized electric and magnetic responses but possesses no electromagnetic coupling. Therefore, the induced moments in such a metasurface are orthogonal: the electric and magnetic dipole moments are directed along the incident electric and magnetic fields, respectively. 

The oscillating moments radiate secondary waves in backward
\begin{equation}
E_{\rm b}=\frac{-j\omega}{2S}\left(\eta_0 p -m\right)
\label{eq:1} \end{equation}
and forward directions
\begin{equation}
E_{\rm f}=\frac{-j\omega}{2S}\left(\eta_0 p +m\right),
\label{eq:2} \end{equation}
where $S=a^2$ is the unit-cell area, $j\omega p/S$ and $j\omega m/S$ are the surface-averaged electric and magnetic currents, respectively; $\eta_0=\sqrt{\mu_0/\epsilon_0}$ is the free space wave impedance. 

Choosing the dipole moments in such a way that the forward radiated secondary wave interfere destructively with the incident one and the backward radiation is absent, 
\begin{equation}
p=\frac{-j S}{\omega \eta_0}E_{\rm inc},\qquad
m=\eta_0 p
\label{eq:3} \end{equation}
one can achieve total absorption in the meta-atom array. Here, $E_{\rm inc}$ is the incident electric field.

Thus, realization of a single-layer meta-atom absorber requires engineering of certain electrically and magnetically polarizable inclusions providing necessary induced dipole moments (\ref{eq:3}).

\section{Realization of a single-layer meta-atom absorber}

\begin{figure}[h!]
\centering
\begin{subfigure}{0.35\columnwidth}
  \centering
  \includegraphics[width=\columnwidth]{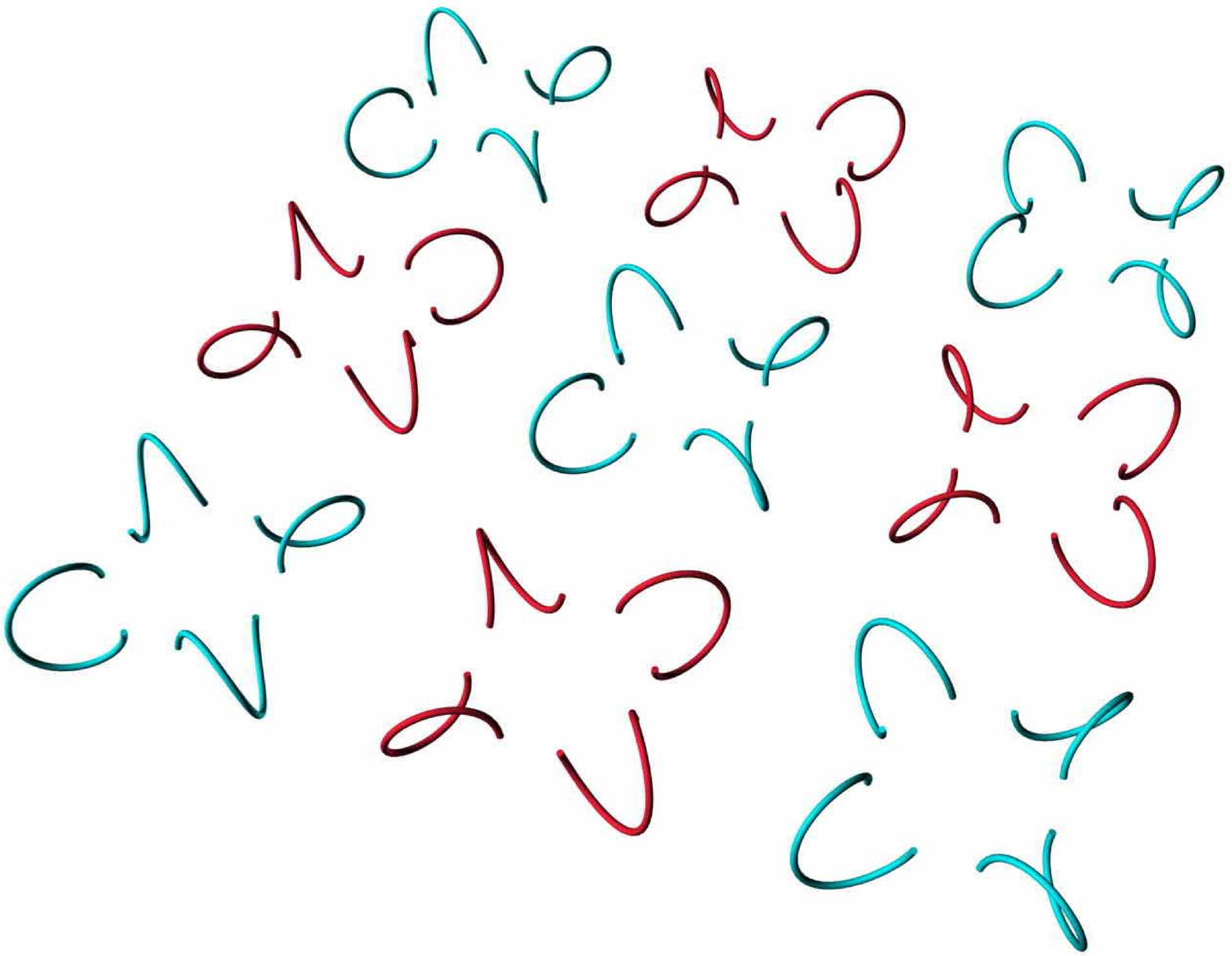}
  \caption{}
  \label{fig:particles}
\end{subfigure}\qquad\qquad
\begin{subfigure}{0.3\columnwidth}
  \centering
  \includegraphics[width=\columnwidth]{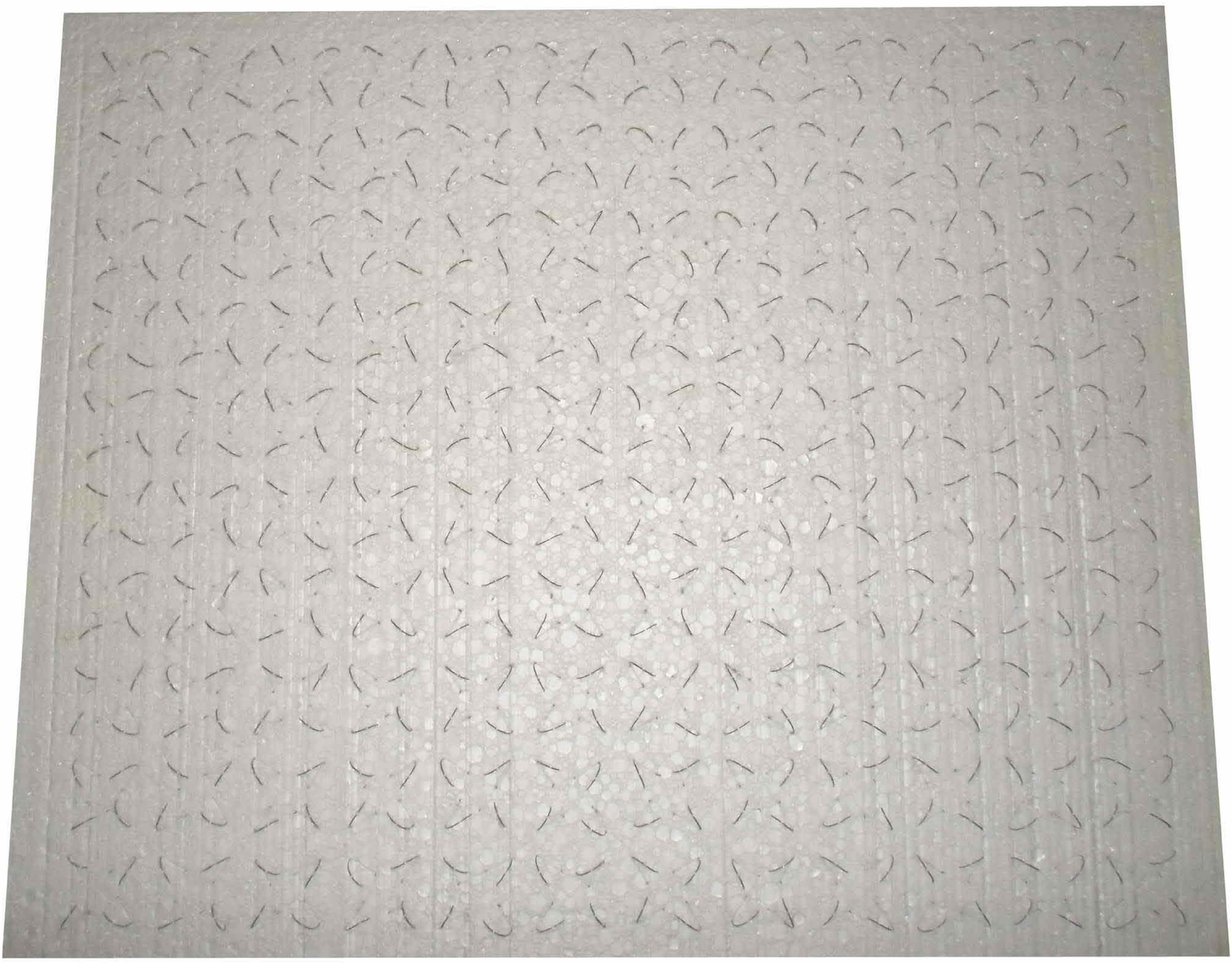}
  \caption{}
  \label{fig:photo}
\end{subfigure}
\caption{(a) A single-layer absorber made of planar array of helical inclusions. Right- and left-handed helices are marked in blue and red, respectively. (b) Experimental sample consisted of paired helical meta-atoms embedded in a plastic foam substrate.}
\label{fig:test2}
\end{figure}

As it was discussed in the previous section, the designed absorbing metasurface must consist of polarizable inclusions with equal electric and magnetic properties. The evident choice is to construct the metasurface from inclusions of two kinds: electric straight wires and magnetic split-ring-resonators. But such a choice is difficult to realize since the inclusions of the different kinds resonate at different frequencies and the inevitable double-resonance performance will destructively affect the absorption level of the structure. Another disadvantage is a thickness which in this case is compared to the wavelength. To avoid these difficulties, we realize a single-layer absorbing metasurface with the use of paired smooth helical inclusions (see Fig.~\ref{fig:particles}) with compensated chiral electromagnetic coupling (it is one of the conditions for total absorption in the metasurface).
The helical inclusions are simultaneously electrically and magnetically polarizable and resonate at the same frequency. An electrically small size ($\lambda$/7) and simple for fabrication shape make them preferable for the use as meta-atoms in the absorber. Compensation of chirality in the metasurface is achieved utilizing right- and left-handed helical inclusions (shown in Fig.~\ref{fig:particles} in blue and red, respectively). The certain arrangement of the helices in the array was considered in \cite{3}.

Equal electric and magnetic responses can be accomplished in the helical inclusions with equal electric and magnetic polarizabilities. Such balanced electromagnetic properties are achieved in a helical inclusion with the following dimensions \cite{4}: the total length of the wire is 46.7~mm, the radius of the helix is 7.2~mm, the height is 11.3~mm, the diameter of the wire is 0.5~mm, the pitch angle of the helix is $14^{\circ}$. The material of the inclusions is nichrome $Ch_{15}Ni_{60}$ with the conductivity $10^6$~S/m. The helical inclusions are embedded in a plastic foam substrate with $\epsilon=1.03$ and thickness 14.4~mm. The arrangement of the helices is illustrated in Fig.~\ref{fig:particles}. Each unit cell consists of four helices of specific handedness located at a distance 19.8~mm from the center of the cell. ``Right-'' and ``left-handed'' unit cells are alternated in the plane of the metasurface with the period 105~mm. 

\begin{figure}[h!]
\centering
\vspace{-1.5cm}
  \includegraphics[width=\columnwidth]{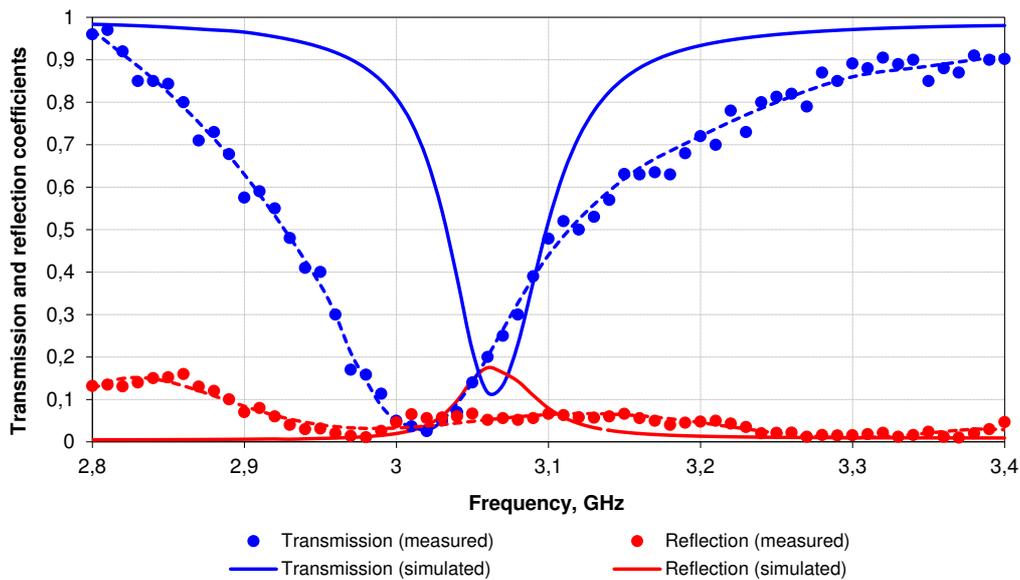}
  \vspace{-2.5cm}
\caption{Normalized intensity of the reflected and transmitted waves from the absorber (normal incidence).}
\label{fig:2}
\end{figure}

The experimental sample (see Fig.~\ref{fig:photo}) of 5$\lambda \times 6\lambda$ size consists of 480~helices. The measurements were carried out in free space in a chamber with two horn antennas (S~band). The sample was illuminated by a normally incident linearly polarized plane wave. Since the metasurface has uniaxial symmetry in the plane, it operates for arbitrary polarized plane waves. Fig.~\ref{fig:2} shows reflection and transmission coefficients for measured and simulated (full-wave simulator ANSYS HFSS) results.

One can see that at the resonance frequency the designed metasurface absorbs $94\%$ of the impinging power. The absorber symmetrically operates from the two sides. Indeed, further increase of the absorption level can be accomplished with thorough optimization of the inclusions. 

\section{Conclusion}
We realized a new kind of electrically thin absorbers of electromagnetic waves. The absorber operates with $94\%$ efficiency at around 3~GHz. The simulated and experimental results are in a good agreement. At the next step, the idea of a single-layer meta-atom absorber can be extended to higher frequencies opening up a possibility to create film absorbers with thickness of one molecular layer.



\end{document}